\documentclass[twocolumn,aps,pre,reprint,groupedaddress]{revtex4}
\usepackage{hyperref, url}

\usepackage{amsmath, amssymb, graphicx, xcolor}
\usepackage{subfigure}

\begin{document}


\title{Effect of money heterogeneity on resource dependency in complex networks}


\author{Harshit Agrawal}
\affiliation{Studio Sirah, Bengaluru 560037, India}
\author{Ashwin Lahorkar}
\affiliation{Pitney Bowes India Pvt. Ltd, Pune 411014, India}
\author{Snehal M. Shekatkar}
\email[]{snehal@inferred.in}
\affiliation{Department of Scientific Computing, Modeling and Simulation, \\S.P. Pune University, Pune 411007, India}



\begin{abstract}
Exchange of resources among individual components of a system is fundamental to systems like a social network of humans and a network of cities and villages. For various reasons, the human society has come up with the notion of \emph{money} as a proxy for the resources. Here we extend the model of resource dependencies in networks that was recently proposed by one of us, by incorporating the concept of money so that the vertices of a network can sell and buy required resources among themselves. We simulate the model using the configuration model as a substrate for homogeneous as well as heterogeneous degree distributions and using various exchange strategies. We show that a moderate amount of initial heterogeneity in the money on the vertices can significantly improve the survivability of Scale-free networks but not that of homogeneous networks like the Erd{\H o}s-R{\'e}nyi network. Our work is a step towards understanding the effect of presence of  money on the resource distribution dynamics in complex networks. 
\end{abstract}


\maketitle

\section{Introduction}
Complex networks have become an indispensable tool to study complex systems from social, biological, and technological domains that are made up of a large number of components interacting with each other in an intricate fashion \cite{newman2018networks,amaral2004complex,boccaletti2006complex,albert2002statistical}. A \emph{dependency-network} is a special type of network in which network edges depict some form of dependency between the vertices of the network. A large number of studies have explored how such dependency can lead to death of a vertex under various conditions \cite{goltsev2006k}. In the last decade or two, there has been a flurry of activity on the problem of dependency between two or more interacting networks \cite{gao2012networks,amini2019sustainable}. 

A particular type of dependency that is common to social systems is the dependency for resources like food. A threshold model of such resource dependencies in networks was proposed in \cite{Ingale_Shekatkar2020} in which a vertex dies if at any time the total amount of resource on it goes below a certain value. This total amount is the sum of the amount produced by the vertex and the amount borrowed from the neighbours that are alive at that time. It was shown there that although \textit{Scale-free} networks tend to survive for substantially longer time than those with homogeneous degree distribution, the latter are actually better because the average vertex survival time for them greatly exceeds that of the former. 

In the real-world, the vertices of a network, for example humans in a social network or cities in a transportation network, usually do not offer surplus resources for free but rather sell those to the ones who need them. This motivates us to ask how does incorporating money in a dependency-network's dynamics affects its behavior in terms of survivability. To this end, we propose a model, which we call the \emph{Request-Offer model}, that uses requests and offers of a resource to give rise to transactions between vertices, where each transaction results into an exchange of certain amount of resource for certain amount of money. After that, we describe results of computer simulations for homogeneous and heterogeneous initial distribution of money on \emph{Scale-free} and \emph{Erd{\H o}s-R{\'e}nyi} networks, and also provide explanation of the observed results before concluding the paper. 

\section{\label{model} Request-Offer model of resource dependency}
Now we describe a model motivated from the model of resource dependency in networks proposed in \cite{Ingale_Shekatkar2020}. The main difference between the two models is that the present model contains \emph{money} that is used by vertices to buy and sell resource from the neighbours. Consider an undirected network with $n$ vertices in which each vertex can either be alive or dead, and initially every vertex is in the alive state. Each vertex $i$ initially has a certain amount of money $M_i$. 
    \begin{enumerate}
        \item{At each discrete time step $t$, each vertex $i$ in the network produces resource amount $X_i(t)$ which is a random variable with a given probability distribution $p(x, \beta_i)$ where $\beta_i$ represents the set of parameters of the distribution. }
        \item{Each vertex $i$ also has a threshold $R_i$ which represents the amount of resource that it needs to survive at each time step. Thus, if the resource amount $X_i(t) < R_i$, the vertex is said to be \emph{in-deficit}, and must try to buy the difference $R_i-X_i(t)$ from its neighbours using the money it has. The vertices with $X_i(t) > R_i$ are said to be \emph{with-surplus}.}
        \item{For simplicity, let's assume that a vertex needs one unit of money to buy one unit of resource. The \emph{in-deficit} vertices, if they have enough money with them, send requests to all their alive neighbours asking for the resource in exchange for money. }
        \item{Each alive vertex thus may receive several requests from the neighbours at each time step, and must decide which ones to respond to. Obviously, if the vertex is itself \emph{in-deficit}, it cannot respond to any requests from the neighbours. If that is not the case, it can respond to those neighbours whose requested amounts are less than or equal to the amount of surplus the vertex has. The vertex \emph{sorts} these `eligible neighbours' using one of several strategies described after the model description, and starts sending offers to them in the order in which they are sorted. Each time an offer is sent, the amount of available surplus reduces (or rather gets reserved) by that much amount independent of whether the neighbour will accept the offer or not. The offers are sent until enough surplus is available for the next offer.}
        \item{Thus, a vertex that has sent out requests in general receives several offers from the neighbours, and accepts one of the offers uniformly randomly. The amount of money equal to the resource received is then transferred to the neighbour whose offer is accepted. This completes the transaction between two vertices.}
\end{enumerate}

    \begin{figure} 
        \centering
        \includegraphics[width=0.95\columnwidth]{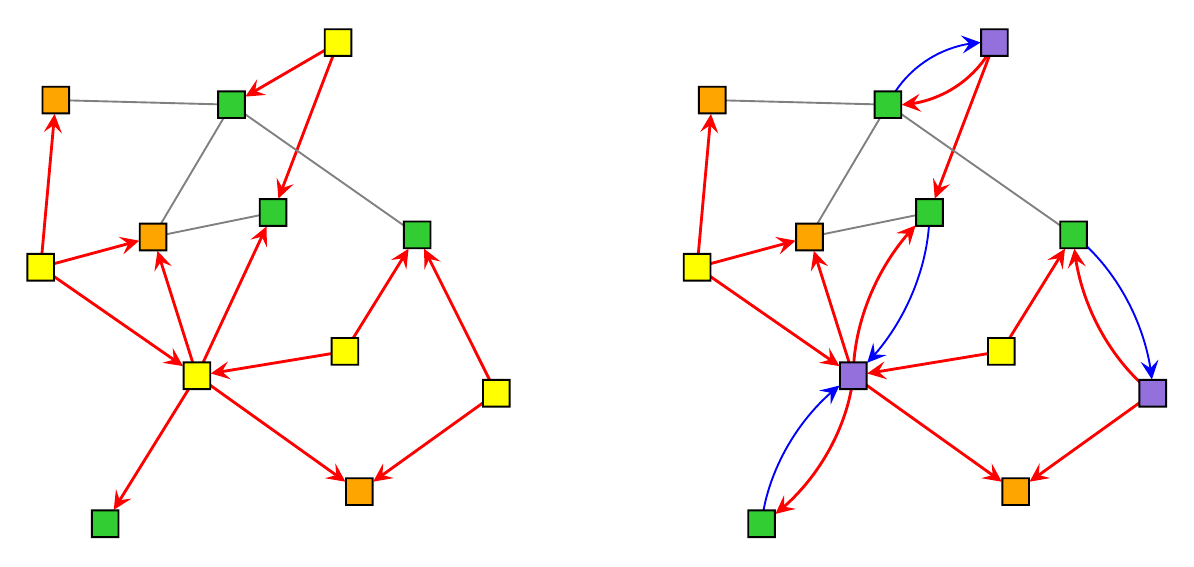}
        \caption{\label{fig:reqoff_graphic} Graphical representation of the Request-Offer model. Left: the \emph{with-surplus} vertices (Green) receive requests (shown as Red arrows) from the \emph{in-deficit} vertices that have enough money (Yellow). The \emph{in-deficit} vertices that do not have enough money (Orange) immediately die. Right: some requests are responded by \emph{with-surplus} vertices (shown as Blue arrows). The \emph{in-deficit} vertices which receive at least one offer, and hence survive, are shown in Purple.}
    \end{figure}
Although the model may look complicated at first sight, it is straightforward since vertices just send requests and offers to each other, and then some offers are accepted leading to a transaction. Fig.~\ref{fig:reqoff_graphic} depicts a compact visual representation of the model.  The only complicated part of the model is where an in-surplus vertex needs to decide the order of the eligible neighbours to send the offers. In real-world, for example in the international trade network of countries, such decisions may depend on lots of factors (e.g. the past trading history and how friendly two given countries are with each other). This motivates us to make use of several \emph{strategies} to sort the eligible neighbours. Some of these are described below:
    \begin{itemize}
        \item{{\bf Random:} Perhaps the most obvious strategy is to order the eligible neighbours in a random order, and then start sending offers to them one by one.}
        \item{{\bf High\_to\_Low:} The eligible neighbours can be ranked in descending order of the amounts they requested. This way, the first offer is always made to the neighbour that has requested the highest amount. }
        \item{{\bf Prop\_to\_Req:} Choose randomly and without replacement among the eligible neighbours, such that the probability of choosing a vertex is proportional to the amount it requested. This is somewhat similar to the High-to-Low strategy except that because of the randomness, the ordering is not exactly descending.}
        \item{{\bf Prop\_to\_Req\_Deg:} Choose randomly and without replacement among the eligible neighbours, such that the probability of choosing the neighbour $j$ is proportional to its requirement $R_j-X_j(t)$ and inversely proportional to $k_j^{\eta}(t)$ where $\eta$ is a \emph{degree bias} parameter. Throughout this paper we fix $\eta=0.6$. This way, the preference is given to the neighbours in high need, but also to those neighbours which have low degree at a given time \cite{Ingale_Shekatkar2020}. }
    \end{itemize}

Here it is worth noting that one of the two important assumptions used in the threshold model in \cite{Ingale_Shekatkar2020} was that the individual vertices do not have the information about the amounts of resource produced on any other vertices. However, as the model description given above makes clear, incorporating money in the model, depending upon the strategy used, may force us to relax this assumption. 
Our interest is mainly in seeing how the survivability of a network depends on its structure and various offer strategies described above. In \cite{Ingale_Shekatkar2020}, two measures of the network survivability were discussed: $\langle T_v\rangle$, the average time for which a vertex in the network survives, and $T_{\text{max}}$ which is the total time for which the network survives. Mathematically, these can be defined as follows:
\begin{equation}
\begin{aligned}
\langle T_v \rangle &= \frac{1}{n}\sum\limits_{v=1}^n T_v \\ 
T_{\text{max}} &= \max\{T_1, T_2, \dots, T_n\}
\end{aligned}
\end{equation}

Here $T_v, v=1, 2, \dots, n$ represent the times for which the individual vertices survive. Note that even when only a single vertex of the network is alive, the network is said to be alive. It was further discussed in \cite{Ingale_Shekatkar2020} that $\langle T_v\rangle$ is a better quantifier of the network survivability than $T_{\text{max}}$ because the latter can be large even when most of the vertices of the network die quickly and only a set of just few vertices survives for long time. Thus, in this paper, we use $\langle T_v\rangle$ to discuss survivability of a network.

In particular, we are interested in analyzing the effect of the degree distribution of the network on its survivability for the specific model of resource dependency proposed here. The null model of choice for such situation is the \emph{configuration model}, which is a random graph model with a give degree distribution \cite{newman2018networks}. To be completely accurate, the configuration model is actually a random graph model with a given degree sequence. However, when the degree sequence is randomly drawn from a specified degree distribution, in the limit $n\to\infty$ this is equivalent to a random graph model with a given degree distribution \cite{peixoto2012}. It is well known that the real-world networks can be roughly classified into two classes based on their degree distributions. The networks whose degree distribution lacks right skewness are called \emph{Homogeneous networks} because the probability of existence of a vertex with substantially high degree compared to the average degree is practically zero in them. The Erd{\H o}s-R{\'e}nyi graph, because of its Poissonian degree distribution, falls in this category \cite{newman2018networks}. On the other hand, the networks in which the degree distribution is right-skewed, can contain vertices with very high degree compared to the average. These are heterogeneous or \emph{Scale-free} networks \cite{barabasi2009scale}. The Price network is an example of a Scale-free network \cite{price1976general}. 

We construct homogeneous and Scale-free random graphs using the configuration model with degree sequences drawn from the Poisson and the power-law degree distributions respectively. Here the term Scale-free is not restricted only for graphs with pure power-law degree distributions, but rather for a class of graphs with fat-tailed degree distributions, and the pure power-law is just a prototype of such graphs. Thus, in this paper, a Scale-free graph would mean a graph whose degree distribution $p_k = Ck^{-\alpha}$ for any degree $k>k_{\text{min}}$ where $C$ is the normalization constant and $\alpha$ is the scaling index; $p_k = 0$ when $k < k_{\text{min}}$. This way, every vertex in the graph has at least degree $k_{\text{min}}$. The normalization constant $C$ is fixed by the condition:
\begin{equation}
\begin{aligned}
\sum\limits_{k=k_{\text{min}}}^{\infty}p_k = 1 \implies C = \frac{1}{\zeta(\alpha, k_{\text{min}})}
\end{aligned}
\end{equation}
where $\zeta(x, a)$ is the Hurwitz zeta function. All the simulations in this paper are performed using $\alpha=2.2$ and $k_{\text{min}}=2$. Similarly, in this paper the term \emph{Homogeneous network} would mean the Erd{\H o}s-R{\'e}nyi graph network, which as mentioned above, has Poissonian degree distribution $p_k = e^{-\langle k\rangle}\langle k\rangle^k/k!$. To have a right comparison, we construct the homogeneous graphs in such a way that they have the same average degree, $\langle k\rangle \approx 9.36$, as that of the Scale-free graphs with given $\alpha$ and $k_{\text{min}}$. At this point, it is important to note that the configuration model allows self-loops and multi-edges in a generated network. The self-loops do not matter as far as our request-offer model is concerned because no requests are sent out if the produced amount is already greater than the threshold, while if the amount is less than the threshold, the vertex cannot offer anything to itself. When there is a multi-edge between a given pair of vertices, we assume that the probability of sending an offer to the requesting vertex in the pair by the other vertex is proportional to the multiplicity of the edge. 

For the simulations of the model, we will henceforth assume that the amount of resource $x(t)$ produced by each vertex at time $t$ is a random number drawn from the Exponential distribution with probability density given by:
\begin{equation}
\begin{aligned}
p(x, \beta) = \frac{1}{\beta}e^{-x/\beta}
\end{aligned}
\end{equation}

This continuous probability distribution has only a single parameter $\beta$ which also happens to be the mean of the distribution. Hence, it can be thought of as the average amount produced by a vertex or its `\textit{production capacity}'. For simplification, let us assume that every vertex has the same `production capacity' $\beta=2$, and also has the same `survival threshold' $R=1$. However, note that at any given time, different vertices would produce different amounts of resource since the production is assumed to be stochastic. Moreover, as already mentioned above, while using the strategy `\textit{Prop\_to\_Req\_Deg}', the degree bias parameter will always be set to $\eta=0.6$.

\section{\label{simulation_results} Simulation results }
First we consider a situation in which initial amount of money is same for all the vertices, and then we demonstrate how introducing heterogeneity in this initial amount can lead to interesting dynamics in the system. After this we provide a detailed explanation of the observed results for both Scale-free and Erd{\H o}s-R{\'e}nyi networks. A ready-to-use implementation of the Request-Offer model described here is available as a part of the Free software Python package \texttt{dependency-networks} \cite{shekatkar2020dependency-networks}.

\begin{figure} 
    \begin{center}
    \includegraphics[width=0.95\columnwidth]{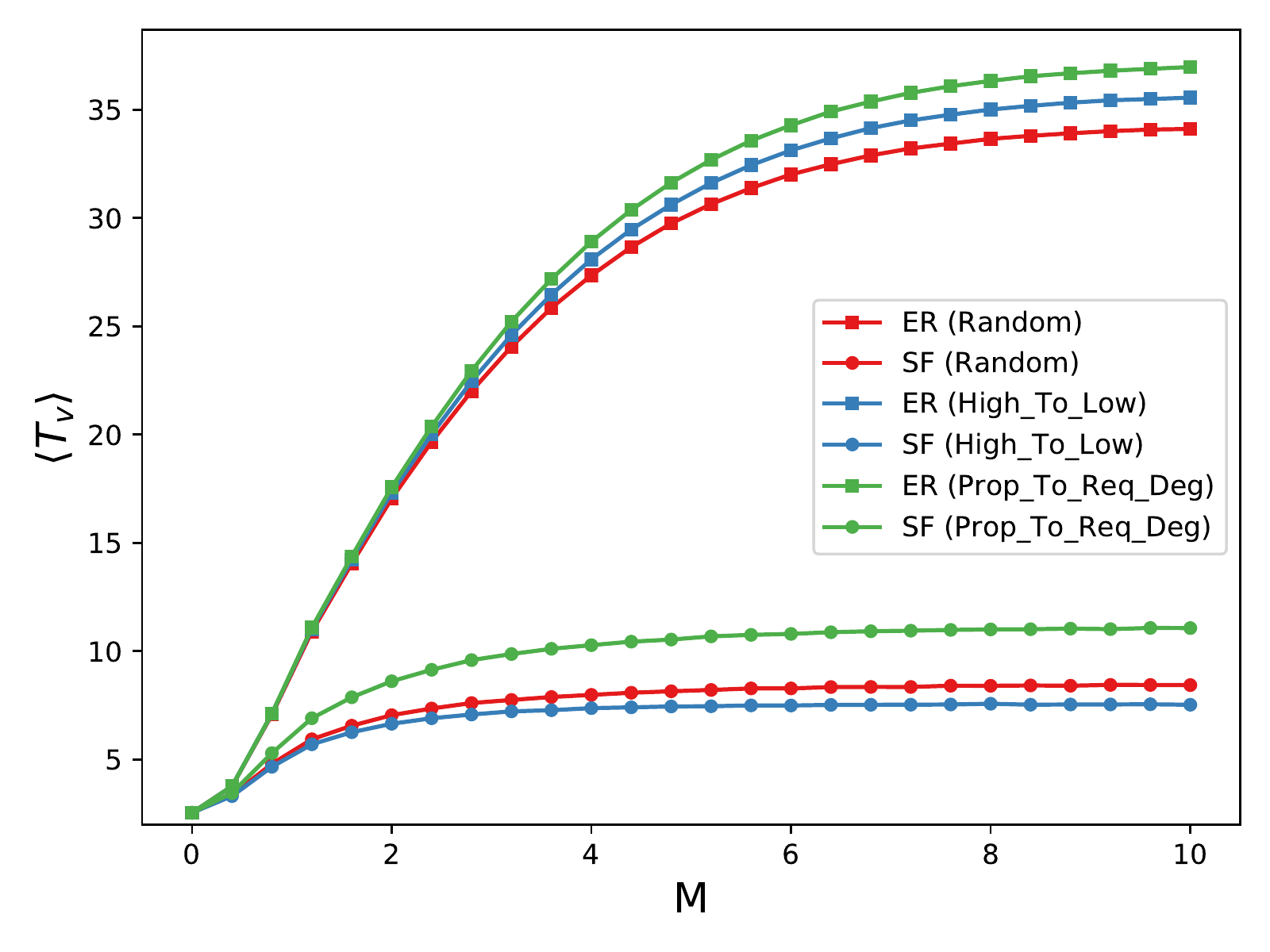}
    \caption{\label{ave_vert_t_vs_money_multi} Variation of $\langle T_v\rangle$ as the initial amount of money $M$ on each vertex is increased for Erd{\H o}s-R{\'e}nyi and Scale-free networks for three different offer strategies. Notice how $\langle T_v\rangle$ saturates for large $M$ independent of the network structure or the offer strategy. The results are averaged over $1000$ random realizations of a network with size $n=10000$. See text for other parameter values.}
    \end{center}
\end{figure}
First let us assume that each of the $n$ vertices in the network has the same fixed amount $M$ of money to start with. Although the amount is the same for each vertex initially, because of the transactions, these amounts would change in time. Fig.~\ref{ave_vert_t_vs_money_multi} shows the variation of the average time for which a vertex in the network survives as a function of the initial amount of money $M$ for Scale-free and Erd{\H o}s-R{\'e}nyi networks. A few things can be noted from the figure. First, $\langle T_v\rangle$ seems to be substantially higher for Erd{\H o}s-R{\'e}nyi networks than for Scale-free ones independent of the offer strategy used, when $M$ is reasonably large. This agrees with the conclusion drawn in \cite{Ingale_Shekatkar2020} that homogeneous networks are more survivable than the heterogeneous ones. Second, when initial money is large enough, increasing it further doesn't lead to increase in the average vertex time. This is understandable because money is just a proxy for resources, and resources in the system are limited. Since increasing money does not increase the actual resources necessary for survival, it cannot increase the survival times for vertices. In other words, being infinitely rich is completely useless in the absence of actual resources. All that money does is to facilitate more and more transactions when surplus resources are available. Thus, the interesting regime for us is the one with low $M$ where limited availability of both money and the surplus resources drives the collective dynamics of the network. Because of this, all further simulations are performed with $M=1$. As already stated before, we assume that one unit of money can buy one unit of resource. 

We also see that different offer strategies lead to different $\langle T_v\rangle$, and the larger the initial money, the larger the difference. The fact that the offer strategy `\textit{Prop\_to\_Req\_Deg}' performs best is understandable from the argument given in \cite{Ingale_Shekatkar2020} because this strategy is just a modified version of the biased distribution strategy presented in the original model. Since high-degree vertices have many neighbours, they have higher chance of receiving at least one offer than lower degree vertices. Hence, somewhat reducing the probability of sending offers to high-degree vertices doesn't affect them much, but at the same time improves the survivability of low-degree vertices which now receive a higher share of the surplus. This increases the overall average vertex time of the network. It is also interesting to note that for the Scale-free network, `\textit{Random}' strategy produces slightly better result than `\textit{High\_to\_Low}', whereas the opposite is true for the Erd{\H o}s-R{\'e}nyi network. 

\begin{figure}[ht]
    \centering
    \includegraphics[width=0.95\columnwidth]{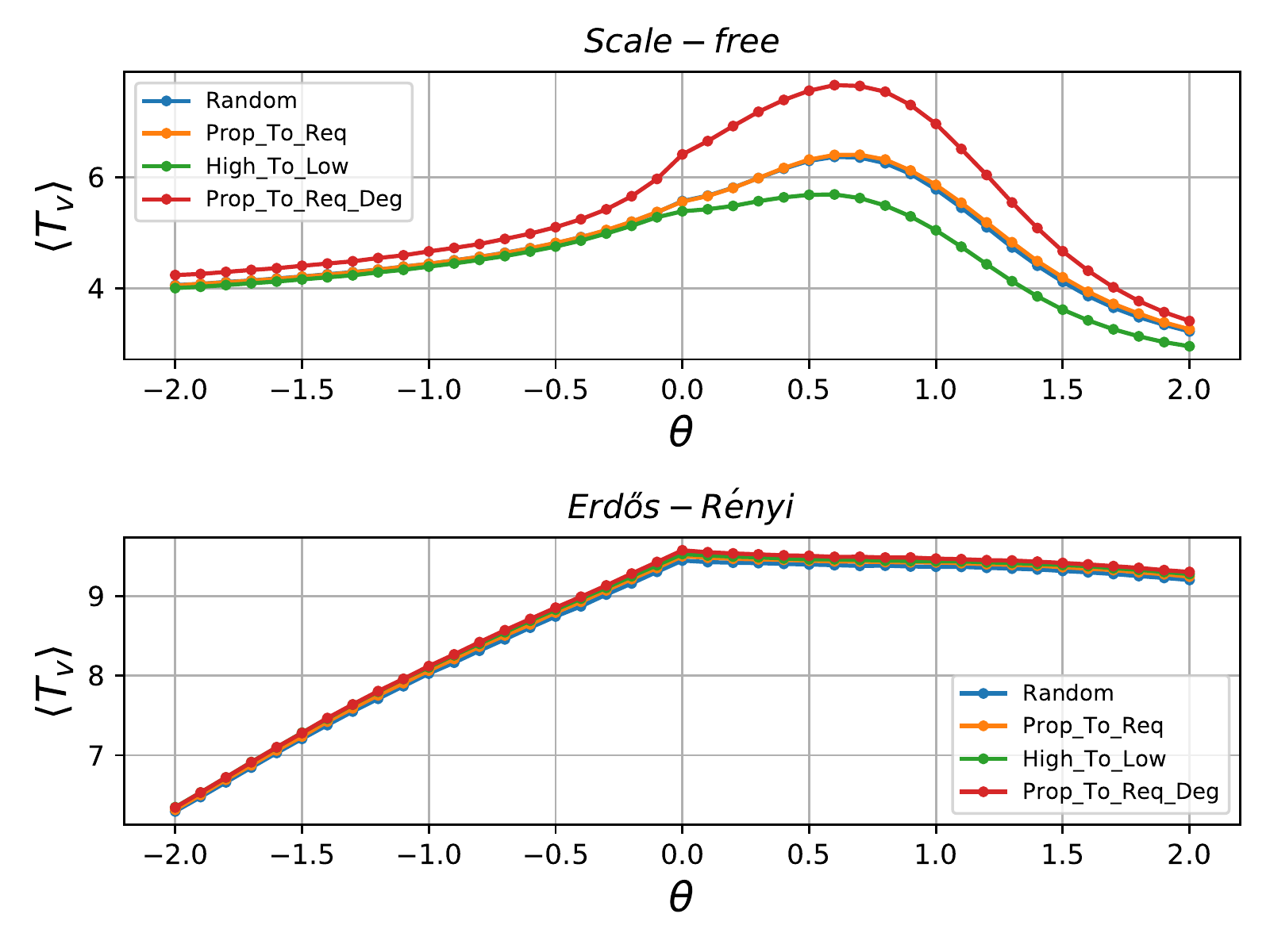}
    \caption{\label{ave_vert_t_vs_theta_money1.0} Variation of $\langle T_v\rangle$ with respect to the money heterogeneity parameter $\theta$ for Erd{\H o}s-R{\'e}nyi and Scale-free networks when $M=1$. The results are averaged over $1000$ random realizations of networks each of size $n=10^4$. See text for discussion. }
\end{figure}
Now we turn to the main problem of checking the effect of initial heterogeneity  in the money on the network survivability. To this end, let us assume that the initial amount of money on a vertex with degree $k$ is proportional to $k^{\theta}$ for some real number $\theta$. The results of the previous section correspond to the special case $\theta=0$. If $\theta > 0$, the high degree vertices would have higher amounts of money initially, whereas, if $\theta < 0$, the opposite would be true. For the homogeneous initial conditions, each vertex initially has money $M$, and hence the total amount of money in the network is $nM$ where $n$ is the size of the network. For a right comparison, with non-zero $\theta$ also we keep the total amount of money to be the same. Thus, we must have: 
\begin{equation}
C\sum\limits_{i=1}^n k_i^{\theta} = nM
\end{equation}
where $C$ is the normalization constant. Thus, initially, the amount of money on vertex $i$ is given by: 
\begin{equation}
M_i(0) = \frac{k_i^{\theta}nM}{\sum\limits_{i=1}^n k_i^{\theta}}
\end{equation}
Rewriting this equation for $M$, we get:
\begin{equation}
M = \frac{1}{n}\sum\limits_{i=1}^n M_i(0)
\end{equation}

Hence, $M$ is the average amount of money on a vertex initially, and can be treated as a parameter just like the case of homogeneous initial distribution. As $\theta$ is varied for a given $M$, the total initial amount in the network remains fixed at $nM$, but its distribution among the vertices changes depending upon their degrees. 

The Fig.~\ref{ave_vert_t_vs_theta_money1.0} shows the plot of average vertex time as a function of $\theta$ for $M=1$ for the Scale-free and the Erd{\H o}s-R{\'e}nyi networks for different offer strategies. An immediate thing to notice from the figure is that the two types of networks behave in radically different ways as the money heterogeneity parameter $\theta$ is varied. For Scale-free networks (top panel), as the value of $\theta$ is increased from negative values, $\langle T_v\rangle$ increases in a non-linear fashion, reaches a maximum value for $\theta \approx 0.8$, and then declines again. As the bottom panel of the figure shows, for the Erd{\H o}s-R{\'e}nyi network, $\langle T_v\rangle$ first linearly increases with $\theta$ and reaches maximum at $\theta=0$, the value at which every vertex has initially the same amount of money, after which it starts decreasing but with a much slower rate than it increased. We should also note that $\theta_{\text{max}}$, the value of $\theta$ for which $\langle T_v\rangle$ is maximized, is almost independent of which offer strategy is in use for both types of networks. To summarize, money heterogeneity seems to increase the survivability of Scale-free networks but not that of homogeneous networks. 

    \begin{figure} 
        \centering
        \includegraphics[width=0.99\columnwidth]{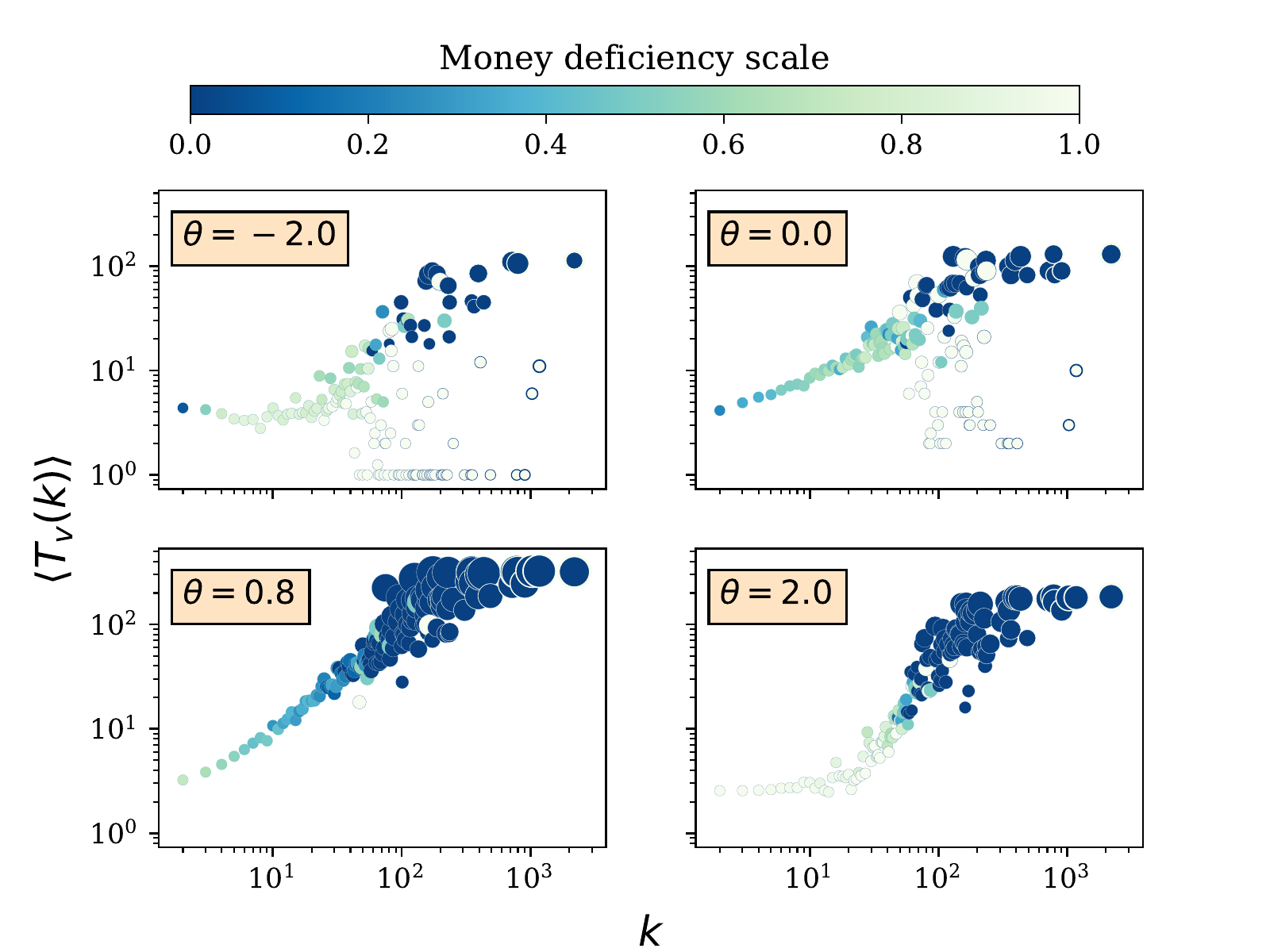}
        \caption{\label{scale_free_deg_time_scatter} Scatter plots of $\langle T_v(k)\rangle$ and degree $k$ for a \textit{Scale-free} network for various values of money heterogeneity parameter $\theta$ with $M=1$ when `\textit{Random}' offer strategy is used. Color scale at the top indicates the value of $\langle \mu(k)\rangle$ (See Eq(\ref{eq:mu})) while the sizes of the markers indicate the value $\langle s(k)\rangle$ (See Eq(\ref{eq:s})). See text for the explanation.}
    \end{figure}

The fact that the average vertex time increases with $\theta$, attains maximum, and then again decreases for Scale-free networks, can be understood as follows. At any given time, an in-deficit vertex can survive only if both of the following conditions are satisfied:
    \begin{enumerate}
        \item{The vertex should have enough money to send out requests to its neighbours.}
        \item{Given that it has sent requests, at least one neighbour should send it an offer.}
    \end{enumerate}

For a low-degree vertex, even if the first of these conditions is met, compared to a high-degree vertex it is harder to meet the second condition since the number of neighbours that could respond is itself low. The situation for a high-degree vertex is different though since if it manages to send out requests, owing to the fact that it has many neighbours, chances that at least one of them would respond are higher. In other words, initially putting a lot of money on low-degree vertices is unproductive because even though they can send out requests, most of them would anyway die because of lack of offers from neighbours. This shows that diverting some part of initial money from low-degree to high-degree vertices should result into survival of more high-degree vertices leading to increase in $\langle T_v\rangle$.

To corroborate the argument presented above, let us define the quantity $\langle T_v(k)\rangle$ as the average time for which a vertex with degree $k$ survives:

\begin{equation}
\label{eq:Tv}
\begin{aligned}
\langle T_v(k)\rangle = \frac{1}{n_k}\sum\limits_{v=1}^n T(v)\delta(k_v, k)
\end{aligned}
\end{equation}

where $T(v)$ is the time for which the vertex $v$ survives, $k_v$ is its degree, $n_k$ is the total number of vertices in the network with degree $k$, and $\delta(i, j)$ is the Kronecker delta which is defined to be $1$ if and only if $i=j$ and $0$ otherwise. We also define another quantity $\langle \mu(k)\rangle$ as:
\begin{equation}
\label{eq:mu}
\begin{aligned}
\langle \mu(k)\rangle = \frac{1}{n_k}\sum\limits_{v=1}^n \mu(v)\delta(k_v, k)
\end{aligned}
\end{equation}
where $\mu(v)=1$ if and only if the vertex $v$ dies because of lack of money and $0$ otherwise. Thus, $\langle \mu(k)\rangle$ measures the fraction of vertices of degree $k$ that die because of lack of money. Finally, we define the quantity $\langle s(k)\rangle$ as:
\begin{equation}
\label{eq:s}
\begin{aligned}
\langle s(k)\rangle = \frac{1}{n_k}\sum\limits_{v=1}^n s(v)\delta(k_v, k)
\end{aligned}
\end{equation}
where $s(v)$ is the number of times vertex $v$ successfully procured required amount of resource from its neighbours when it was in-deficit during its lifetime. Thus, the quantity $\langle s(k)\rangle$ is the expected number of times a vertex with degree $k$ is saved by its neighbours during its lifetime. 

Fig.~\ref{scale_free_deg_time_scatter} shows $\langle T_v(k)\rangle$ vs $k$ scatterplots for four different values of the money heterogeneity parameter $\theta$ for one \textit{Scale-free} graph when `\textit{Random}' offer strategy is used. The sizes of the markers in the plots indicate $\langle s(k)\rangle$ while the color scale indicates $\langle \mu(k)\rangle$. To discuss these scatterplots in detail, we will need some quantitative definition of \textit{low-degree} and \textit{high-degree} vertices. If $\langle k\rangle$ denotes the average degree of the network, it is reasonable to call the vertices with $k < \langle k\rangle$ as \textit{low-degree} while those with $k > \langle k\rangle$ as \textit{high-degree} vertices. Although this definition is somewhat arbitrary, as we will see, it will serve the purpose of explaining the variation of $\langle T_v\rangle$ with $\theta$. In our simulations, we have used graphs with $\langle k\rangle \approx 9.36$, and hence for us all the vertices with degree $9$ or less are the low-degree ones.

We can now rewrite $\langle T_v\rangle$ in terms of the networks degree distribution $p_k$ and $\langle T_v(k)\rangle$ as:
\begin{equation}
\label{eq:Tv_HL}
\begin{aligned}
\langle T_v\rangle &= \sum\limits_{k=k_{\text{min}}}^{\infty} p_k \langle T_v(k)\rangle\\ 
  &= \sum\limits_{k=k_{\text{min}}}^{\lfloor \langle k\rangle\rfloor} p_k \langle T_v(k)\rangle + \sum\limits_{\lceil \langle k\rangle\rceil}^{\infty} p_k\langle T_v(k)\rangle,\\
 \therefore \langle T_v\rangle  &= f_{L}\langle T_v\rangle_{\text{low}} + f_{H}\langle T_v\rangle_{\text{high}}
\end{aligned}
\end{equation}

where $\langle T_v\rangle_{\text{low}}$ and $\langle T_v\rangle_{\text{high}}$ are the average vertex times for low-degree and high-degree vertices respectively, and $f_L$ and $f_H$ are the fractions of low-degree and high-degree vertices in the network:
\begin{equation}
\begin{aligned}
f_L = \sum\limits_{k=k_{\text{min}}}^{\lfloor \langle k\rangle\rfloor} p_k\quad \text{and} \quad f_H = \sum\limits_{\lceil \langle k\rangle\rceil}^{\infty} p_k
\end{aligned}
\end{equation}

It is important to note that for Scale-free networks, $f_L$ is much higher than $f_H$. With our choice of parameters $k_{\text{min}} = 2$ and $\alpha=2.2$, we have:
\begin{equation}
\begin{aligned}
f_H &= \sum\limits_{\lceil \langle k\rangle\rceil}^{\infty} p_k = \frac{1}{\zeta(\alpha, k_{\text{min}})}\sum\limits_{\lceil \langle k\rangle\rceil}^{\infty} k^{-\alpha} \approx 0.11 \\
f_L &= 1-f_H \approx 0.91
\end{aligned}
\end{equation}
That is, more than $90\%$ vertices in our \textit{Scale-free} network are low-degree vertices. This directly implies from Eq(\ref{eq:Tv_HL}) that unless the average vertex time of high-degree vertices $\langle T_v\rangle_{\text{high}}$ is substantially larger than that of low-degree vertices, network $\langle T_v\rangle$ is mostly determined by $\langle T_v\rangle_{\text{low}}$. We are now in a position to discuss the scatterplots in Fig.~\ref{scale_free_deg_time_scatter}. 

First let us look at the plot for $\theta=0$ case (top right) from Fig.~\ref{scale_free_deg_time_scatter}, corresponding to homogeneous initial distribution of the money. The plot indeed shows that the average time for which a vertex survives tends to increase with its degree. However, the plot also shows a significant number of \textit{high-degree} vertices with quite low lifetimes, sometimes even lower than the \textit{low-degree} ones. We have seen above that a high-degree vertex that is in-deficit has a higher chance of getting at least one offer from its neighbours than a low-degree vertex provided that it is able to send out requests in the first place. But if a vertex does not have enough money, it can't send requests to its neighbours and hence can't get any offer even if its degree is high. The $\theta=0$ panel clearly shows that the \textit{high-degree} vertices with lower lifetimes are precisely those which lacked money (Whitish circles). The plot also shows that the average number of times the lowest degree vertices (e.g. $k=2$ and $k=3$) are saved by their neighbours is small as indicated by sizes of the corresponding markers. The same markers are also darker which implies that most of them had sufficient money to send out requests. This shows that money that was put on these low-degree vertices was completely wasted since having more money couldn't save them. 

\begin{figure} 
        \centering
        \includegraphics[width=0.8\columnwidth]{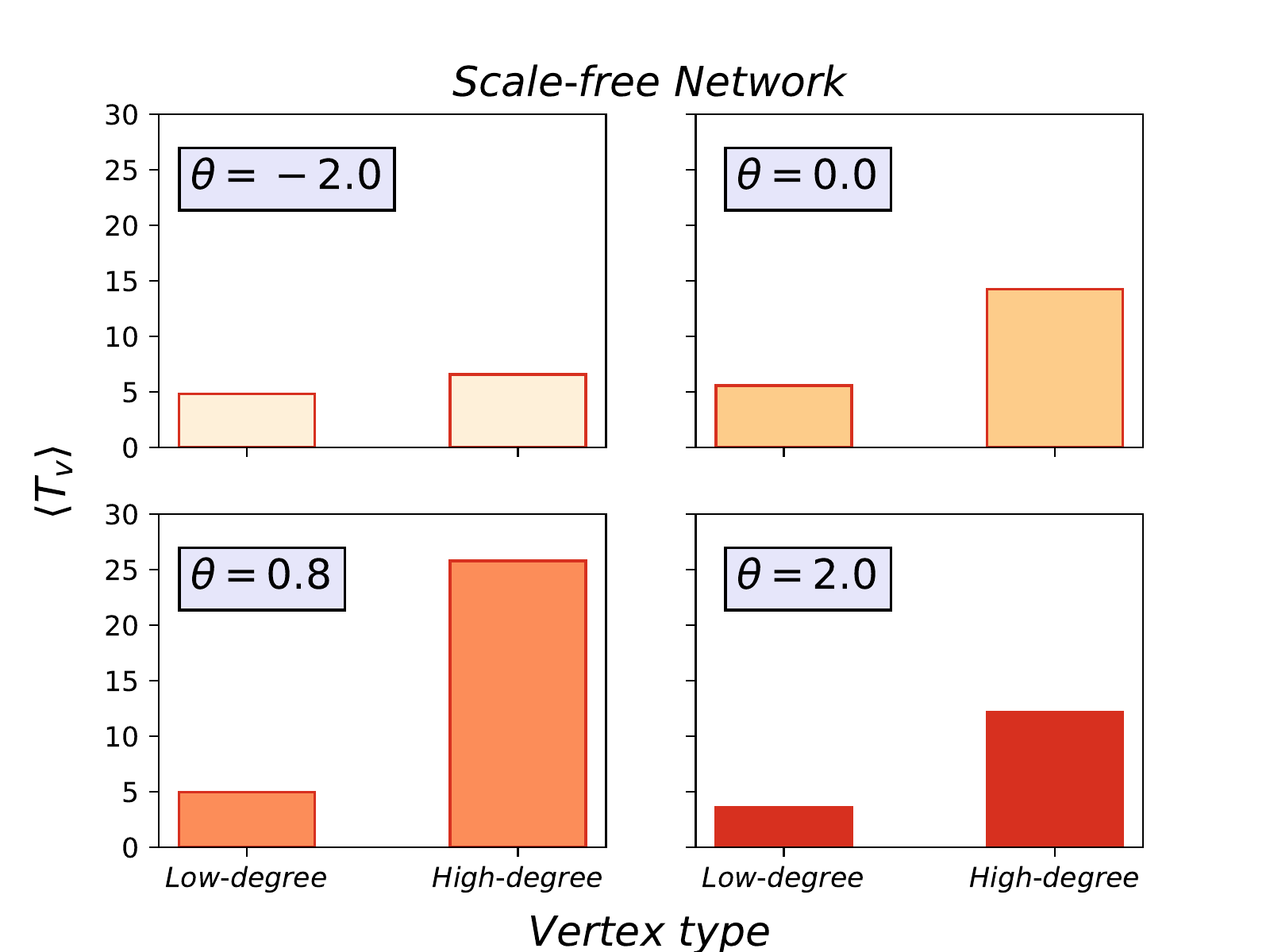}
        \caption{\label{sf_low_high_bar} Average vertex times $\langle T_v\rangle$ for low-degree and high-degree vertices for \textit{Scale-free} network for different values of money heterogeneity parameter $\theta$. Results are averaged over $1000$ random realizations. Error bars are too small to show.}
    \end{figure}

When we divert some of these initial money on low degree vertices to \textit{high-degree} vertices by increasing $\theta$, the situation dramatically changes as seen in the scatterplot for  $\theta=0.8$ (Bottom panel). Now the lowest-degree vertices can be seen to have become lighter in color because they couldn't send requests in the first places when in need. But as anyway they wouldn't have survived had they been able to send out requests, that money is better spent by higher degree vertices. As the plot shows, the high-degree whitish circles have now almost disappeared since now they are able to send out requests and their requests are also responded by their neighbours. This increases the survival time for high-degree vertices without much affecting the average survival time of low-degree vertices, and consequently increases the average vertex time $\langle T_v\rangle$ for the network. We can also explicitly see these times in Fig.~\ref{sf_low_high_bar} for the same $\theta$ values as in Fig.~\ref{scale_free_deg_time_scatter}. As this figure shows, $\langle T_v\rangle$ for low-degree vertices are almost same for $\theta=0.0$ and $\theta=0.8$, but those for high-degree vertices differ substantially. 

The behaviour for extreme cases $\theta=-2.0$ and $\theta=2.0$ can be explained in a similar manner. When $\theta=-2.0$, initially most of the money in the network is put on the low degree vertices. But since most of the time their requests are not answered, their average survival time is not much different than $\langle T_v\rangle_{\text{low}}$ for $\theta=0.0$ or $\theta=0.8$ (See Fig.~\ref{sf_low_high_bar}). But as the scatterplot for $\theta=-2.0$ shows, in this case most of the high-degree vertices die because of lack of money resulting in the reduction of $\langle T_v\rangle_{\text{high}}$ and consequently that of $\langle T_v\rangle$. On the opposite extreme when most of the initial money in the network is put on the high-degree vertices in the network by increasing $\theta$ towards high positive values, as the scatterplot for $\theta=2.0$ shows, survivability of low-degree vertices is appreciably lower compared to other $\theta$ values shown in the plot. We have already discussed that due to high values of $f_L$, the fraction of low-degree vertices in a \textit{Scale-free} network, even a small reduction in $\langle T_v\rangle_{\text{low}}$ can significantly decrease the network $\langle T_v\rangle$. This explains why beyond $\theta=0.8$, we see continuous decrease of $\langle T_v\rangle$ in Fig.~\ref{ave_vert_t_vs_theta_money1.0}.

As Fig.~\ref{ave_vert_t_vs_theta_money1.0} shows, Erd{\H o}s-R{\'e}nyi network clearly behaves differently than the Scale-free network as money heterogeneity is varied. To understand this behaviour, we again use the same analysis that we used for \textit{Scale-free} networks. The main difference between the two types that directly affects their behaviours is the difference between $f_L$ values. We have seen that $f_L$ is very large compared to $f_H$ in a \textit{Scale-free} network. Now let us compute $f_H$ for an Erd{\H o}s-R{\'e}nyi network for which the degree distribution $p_k$ is the Poisson distribution:
\begin{equation}
\begin{aligned}
p_k = e^{-\langle k\rangle}\frac{\langle k\rangle^k}{k!}
\end{aligned}
\end{equation}
Thus, with $\langle k\rangle = 9.36$ we have:
\begin{equation}
\begin{aligned}
f_H &= \sum\limits_{\lceil \langle k\rangle\rceil}^{\infty} p_k = e^{-\langle k\rangle}\sum\limits_{\lceil \langle k\rangle\rceil}^{\infty} \frac{\langle k\rangle^k}{k!} \approx 0.46\\
f_L &= 1-f_H \approx 0.54
\end{aligned}
\end{equation}
    \begin{figure} 
        \centering
        \includegraphics[width=0.8\columnwidth]{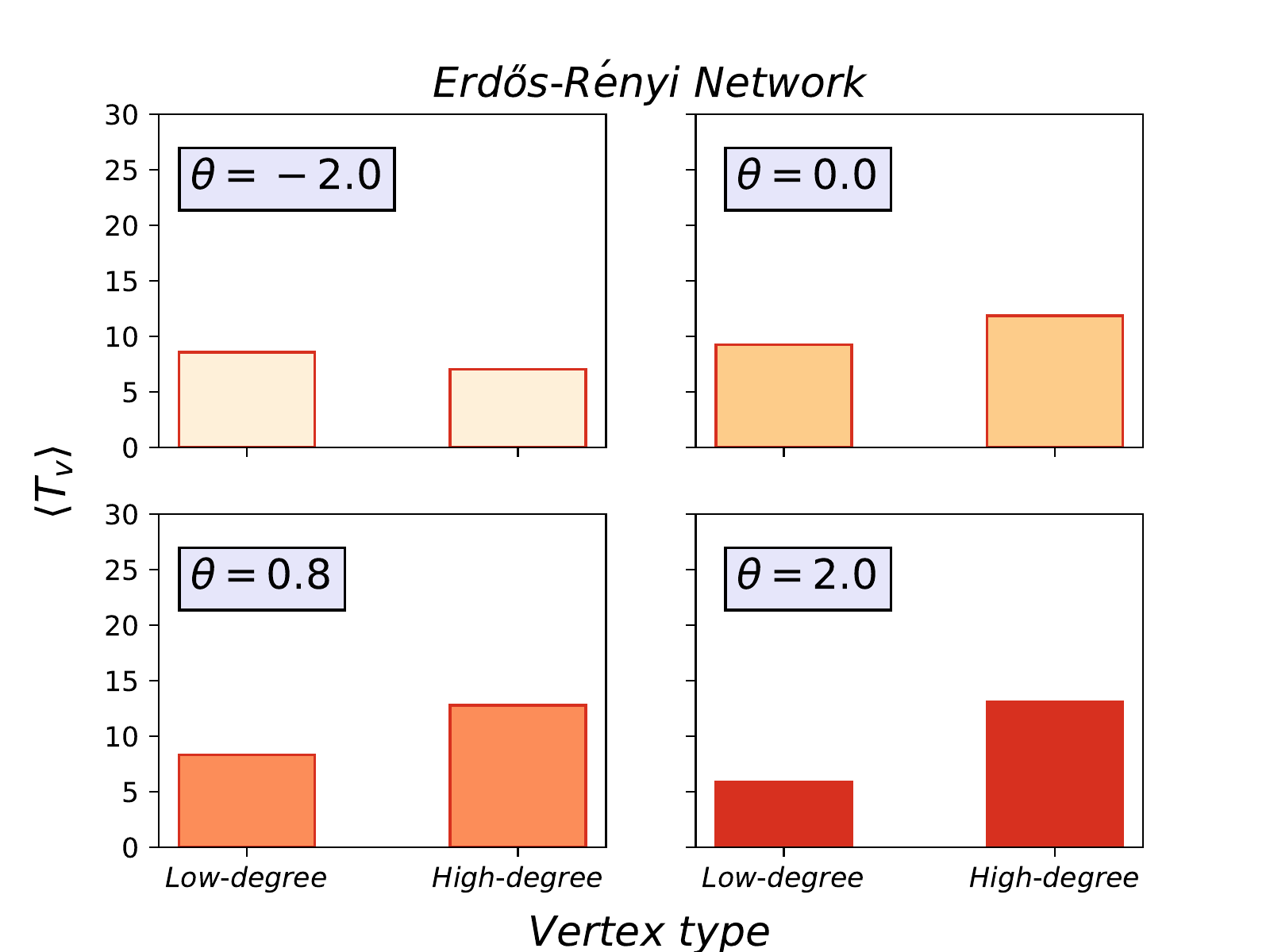}
        \caption{\label{er_low_high_bar} Average vertex times $\langle T_v\rangle$ for low-degree and high-degree vertices for \textit{Erd{\H o}s-R{\'e}nyi} network for different values of money heterogeneity parameter $\theta$. Results are averaged over $1000$ random realizations. Error bars are too small to show.}
    \end{figure}
\noindent Thus, the $\langle T_v\rangle$ of Erd{\H o}s-R{\'e}nyi network is decided almost equally by both low-degree and high-degree vertices. For $\theta=-2.0$, just like the Scale-free case, substantial amount of money resides on low-degree vertices and most high-degree vertices die of lack of money leading to decrease in $\langle T_v\rangle_{\text{high}}$. However, since $f_H$  is quite close to $f_L$, this immediately lowers network $\langle T_v\rangle$ (See Eq(\ref{eq:Tv_HL})). On the other extreme, for $\theta=2.0$, substantial money resides on high-degree vertices making it possible for them to send out requests. Since for them probability of getting at least one offer is also large, this results into significant increase in $\langle T_v\rangle_{\text{high}}$. As is evident from Fig.~\ref{er_low_high_bar}, although lowering money on low-degree vertices does affect them to some extent, since they are anyway bad at procuring offer whether they send out requests or not, decrease in $\langle T_v\rangle_{\textit{low}}$ is not as much as huge gain in $\langle T_v\rangle_{\textit{high}}$. Again, since $f_L$ and $f_H$ are not very different in this case, network $\langle T\rangle$ in this case manages to be only slightly lower than that for $\theta=0.0$.

\section{\label{conclusions}Conclusions}

The central question that we discussed in this paper is the effect of initial heterogeneity in money on network survivability. We showed that a right amount of heterogeneity is exceptionally beneficial for Scale-free networks but not for homogeneous networks like the Erd{\H o}s-R{\'e}nyi network. The main observation that allowed us to explain this difference is that low-degree vertices, even when they have enough money to send out offers, cannot survive much because they don't have enough neighbours who can offer them help when needed.

Our work here provides new insights for understanding the distribution of resources in a networked system when money is present in the system. We hope that these insights would prove useful for effectively distributing resources in real-world networks in which money is an essential component of the process of distribution.
\begin{acknowledgments}
AL would like to thank Pitney Bowes India for support to carry out this research. SMS would like to thank the Department of Science and Technology (DST), India for the financial support in the form of DST-INSPIRE Faculty Fellowship (DST/INSPIRE/04/2018/002664) under which this work was carried out.
\end{acknowledgments}
%
\end{document}